\documentclass[11pt,a4paper]{article}

\usepackage{multirow}
\usepackage{booktabs}
\usepackage{algorithm}
\usepackage{algorithmic}
\usepackage{graphicx}
\usepackage{amssymb,amsmath}
\usepackage[top=2cm, bottom=2cm, outer=2.5cm, inner=2.5cm]{geometry}

\newcommand{\Norm}{\mathcal{N}}
\newcommand{\vect}[1]{\boldsymbol{#1}}

\begin{document}

\title{MCMC inference for Markov Jump Processes via the Linear Noise Approximation}

\author{Vassilios Stathopoulos\footnote{1 v.stathopoulos@ucl.ac.uk} and Mark A. Girolami\footnote{2 m.girolami@ucl.ac.uk}\\
Department of Statistical Science,\\ Centre for Computational Statistics and Machine Learning,\\
University College London, Gower Street, London WC1E 6BT, UK
}


\date{November 2012}

\maketitle

\begin{abstract}
Bayesian analysis for Markov jump processes is a non-trivial and challenging problem. Although exact inference is theoretically possible, it is computationally demanding thus its applicability is limited to a small class of problems. In this paper we describe the application of Riemann manifold MCMC methods using an approximation to the likelihood of the Markov jump process which is valid when the system modelled is near its thermodynamic limit. The proposed approach is both statistically and computationally efficient while the convergence rate and mixing of the chains allows for fast MCMC inference. The methodology is evaluated using numerical simulations on two problems from chemical kinetics and one from systems biology.
\end{abstract}

\section{Introduction}\label{sec:introduction}

Markov Jump Processes (MJP) provides us with a formal description of the underlying stochastic behaviour of many physical systems and as such they have a wide applicability in many scientific fields. In chemistry and biology, for example, they are applied for modelling reactions between chemical species \cite{Gillespie2007,Komorowski2011}. In ecology and epidemiology, they are used for modelling the population of interacting species in the environment \cite{Spencer2005} while in telecommunications they describe the population of information packets over a network \cite{Adas1997}. In order to introduce some terminology and notation we will give a more concrete example from chemical kinetics. However, the modelling methodology is similar in other applications although different assumptions are needed, depending on the system being modelled, for calculating reaction rates. Consider a model for the population of molecules of two interacting chemical species, $X_{A}$ and $X_{B}$, in a solution of volume $\Omega$, where $X_{A}$ and $X_{B}$ denote the number of molecules of chemicals $A$ and $B$ respectively. The interactions between the species are modelled using \emph{reactions} which are specified using the following notation: $R_{1}: A+B \xrightarrow{c_{1}} 2A$. On the left hand side appear the \emph{reactants} and on the right hand side the \emph{products} of the reaction while over the arrow appears the \emph{rate constant} $c_{1}$ which is the probability that a randomly chosen pair of $A$ and $B$ will react according to $R_{1}$. This reaction, for example, specifies that a pair of molecules $A$, $B$ react with probability $c_{1}$ to produce a new molecule of $A$. For calculating the probability of a reaction taking place given the current state of the system, \textit{i.e.} the number of molecules of chemicals $A$ and $B$, several system dependent assumptions must be made. For chemical reactions it is assumed that in a well stirred solution the probability of a reaction is proportional to the populations of its products \cite{Gillespie2005}. For $R_{1}$ we can write it as $f_{1}(X_{A},X_{B},c_{1}) = c_{1}\Omega^{-1}X_{A}X_{B}$. Following the same reasoning additional reactions and species can be added in order to construct large and complex reaction networks. Together, the state of the system $X_{A},X_{B}$, the set of reactions and the reaction rates specify a Markov Jump process where the occurrences of reactions are modelled as a Poisson process.

 For this particular example the probability of the reaction has a simple form and is linear with respect to the populations. However in many real applications this is often not the case while the \emph{rate constants}, $c_{1}$, are unknown. Given a fully specified MJP, \textit{i.e.} a MJP with known parameters, rate constants and initial conditions, it is possible to perform exact simulation and obtain samples from the underlying stochastic process using the Stochastic Simulation Algorithm (SSA) of \cite{Gillespie2007}. In many problems there are system parameters which are not specified or are unknown while it is relatively easy to collect partial observations of the physical process at discrete time points. The interest is therefore to obtain statistical estimates of the unknown parameters using the available data. 

As a consequence of the Markov property, MJPs satisfy the Chapman-Kolmogorov equation from which we can directly obtain the forward master equation describing the evolution of the system's state probability over any time interval. However, even for small and simple systems the master equation is intractable and it is not straightforward as to how partially and  discretely observed data from the physical process should be incorporated in order to perform inference over unknown system parameters. Recently, \cite{Boys2008} have shown that it is possible to construct a Markov Chain whose stationary probability distribution is the posterior of the unknown parameters without resorting to any approximations of the original MJP. Their method however is computationally expensive while the strong correlation between posterior samples means that a large number of MCMC iterations are required in order to obtain Monte Carlo estimates with sufficient accuracy. 

An alternative is to consider suitable approximations of the likelihood function. The system size expansion of \cite[Chap. 10]{vanKampen1992} provides a systematic method for obtaining approximations of a physical process approaching its thermodynamic limit. The most simple approximation yields the Macroscopic Rate Equation (MRE) which describes the thermodynamic limit of the system with a set of ordinary differential equations neglecting any random fluctuations. Although the MRE has been extensively studied in the literature, see for example, \cite{Xu2010,Calderhead2011}, it is not applicable for problems where information about the noise and the random fluctuations is necessary or the system is far from its thermodynamic limit. The diffusion approximation \cite{vanKampen1982,Gillespie2000} describes the physical process by a set of non-linear stochastic differential equations with state dependent Brownian motion. Similar to the master equation however, the likelihood is intractable. In \cite{Roberts2001} a transformation is applied such that the Brownian increments are independent of the system state and thus the system can be easily simulated. However this limits the applicability of the methodology into systems where such a transformation is possible. A more general methodology in presented in \cite{Golightly2011} where an approximation of the likelihood is used instead. Finally, a less studied approach for the purpose of inference is the Linear Noise Approximation (LNA) which conveniently decouples non-linearity in the diffusion approximation into a non-linear set of ordinary differential equations in the MRE and a set of linear stochastic differential equations for the random fluctuations around a deterministic state \cite[Chap. 10]{vanKampen1992},\cite{Wallace2012}. Recently, \cite{Komorowski2009} have shown the simple analytic form of the approximate likelihood obtained by the LNA simplifies MCMC inference and can be applied to problems with relatively small number of molecules.

A commonly employed algorithm for MCMC is the Metropolis-Hastings algorithm \cite{Robert2005}, which relies on random perturbations around the current state using a local proposal mechanism. It should be noted here that the state of the Markov chain is different from the state of the stochastic process. In the MCMC context state refers the current values of the unknown system parameters whereas the state of the system refers to the value of the stochastic process at a given time. We will use the term state interchangeably for the rest of this paper and its meaning will be clear from the context. Due to the local nature of the proposal mechanism used by the Metropolis-Hastings algorithm, samples from the posterior exhibit strong random walk behaviour and auto-correlation. Tuning the proposal mechanisms to achieve good mixing and fast convergence is far from straightforward even though some theoretical guidance is provided \cite{Roberts1997}. MCMC methods, such as the Metropolis Adjusted Langevin Algorithm (MALA) \cite{Roberts2003} and the Hamiltonian Monte Carlo (HMC) \cite{Duane1987}, have also been studied in the literature and have been shown to be more efficient than random walk Metropolis-Hastings in terms of Effective Sample Size (ESS) and convergence rates on several problems. However, HMC and MALA also require extensive tuning of the proposal mechanisms, see for example \cite{Neal93} and \cite{Roberts1998}. For MJPs the problem is compounded further since system parameters, such as probability rate constants of chemical reactions, are often highly correlated and whose values may differ by orders of magnitudes. The resulting posterior distributions have long narrow ``valleys'' preventing any local proposal mechanism from proposing large moves about the parameter space. 

More recently \cite{Girolami2011} proposed exploitation of the underlying Riemann manifold of probability density functions when defining MCMC methods thus exploiting the intrinsic geometry of statistical models, thereby providing a principled framework and systematic approach to the proposal design process. These algorithms rely on the gradient and Fisher Information matrix of the likelihood function to automatically tune the proposal mechanism such that large moves on the parameter space are possible and therefore improve convergence and mixing of the chains. In \cite{Calderhead2011} this approach has been successfully applied for the MRE approximation of chemical reaction networks. For the LNA the Fisher Information and the gradient of the likelihood function can be easily obtained \cite{Komorowski2011}. In this paper we study the application of the Riemann manifold MCMC methods for the LNA approximation and compare the mixing efficiency and computational cost with to the commonly used Metropolis-Hastings algorithm. Moreover we study how the the Markov chains and the resulting Monte Carlo estimates behave for systems which are far from their thermodynamic limit. The aim is to improve the efficiency of MCMC inference for MJPs in order to allow for larger and more complex models frequently encountered in biology and chemistry to be studied in more detail.

In the next section we give a brief overview of Markov jump processes. The diffusion and linear noise approximations are presented in section \ref{sec:Approximations}. We then discuss MCMC and the Riemann manifold algorithms in section~\ref{sec:MCMC}. Numerical simulations are presented in section~\ref{sec:experiments} while section~\ref{sec:conclusions} concludes the paper.

\section{Markov Jump Processes}
\label{sec:MJP}
A $D$-dimensional stochastic process is a family of $D$ random variables $\vect{X}(t)=[X_{1}(t),\dots,X_{D}(t)]^{T}$ indexed by a continuous time variable $t$ with initial conditions $\vect{X}(t_{0}) = \vect{x}_{t_{0}}$. A Markov Jump Process (MJP) is a stochastic process satisfying the Markov property such that 
$$p[\vect{X}(t_{0}),\dots,\vect{X}(t_{N})] = p[\vect{X}(t_{0})]\prod_{i=1}^{N}p[\vect{X}(t_{i})|\vect{X}(t_{i-1})],$$
where the dependence on any parameters or other quantities has been suppressed.
That is, the conditional probability of the system state at time $t_{i}$ only depends on state of the system at the previous time $t_{i-1}$. A MJP is characterised by a finite number, $M$, of state transitions with rates $f_{j}(\vect{x},\vect{\theta},t)$ and state change vectors $\vect{s}_{j}=(s_{1,j},\dots,s_{D,j})^{T}$ with $j\in[1,\dots,M]$. $f_{j}(\vect{x},\vect{\theta},t)dt$ is the probability, given the state of the system at time $t$, $\vect{X}(t)=\vect{x}$, of a jump to a new state $\vect{x}+\vect{s}_{j}$ in the infinitesimal time interval $[t,t+dt)$. For the problems we consider in this paper the transition rates not only depend on the current state and time but also on unknown rate parameters $\vect{\theta}$. From the Markov property we can directly obtain the conditional probability of the system being in state $\vect{x}$ at time $t$ given initial conditions which is characterised by the master equation
 \begin{equation}
 \frac{p(\vect{x},t|\vect{x}_{0},t_{0})}{dt} = \sum_{j=1}^{M}\left[ f_{j}(\vect{x}-\vect{s}_{j},\vect{\theta},t)p(\vect{x}-\vect{s}_{j},t|\vect{x}_{0},t_{0}) -  f_{j}(\vect{x},\vect{\theta},t)p(\vect{x},t|\vect{x}_{0},t_{0})\right].
 \label{eq:masterEquation}
\end{equation}

Equation~(\ref{eq:masterEquation}) in general form is intractable especially when the transition rate functions $f_{j}(\cdot)$ are nonlinear with respect to the system state. Numerical simulation is also prohibitively expensive as the computational cost grows exponentially with $D$ \cite{Ferm2008}.

However, given initial conditions $\vect{X}(t_{0}) = \vect{x}_{t_{0}}$ and values for the unknown rate parameters $\vect{\theta}$ we can simulate realisations of the MJP by first noting that the time $\tau$ to the next state transition is exponentially distributed with rate $\lambda = \sum_{j=1}^{M}f_{j}(\vect{x}_{t_{0}},\vect{\theta},t_{0})$ and the new state $\vect{X}(t_{0}+\tau)$ will be $\vect{x}_{t_{0}} + \vect{s}_{j}$ with probability
$f_{j}(\vect{x}_{t_{0}},\vect{\theta},t)/\lambda.$
This results in an iterative algorithm from which we can forward simulate a complete trajectory for the stochastic process $\vect{X}(t)$, known as the Stochastic Simulation Algorithm (SSA) \cite{Gillespie2007} in the chemical kinetics literature.

From the specification of the MJP we can also write the likelihood function with respect to the parameters $\vect{\theta}$ for a completely observed process $\vect{X}(t)$ at the time interval $[0,T]$ as 
$$p(\vect{X}|\vect{\theta}) = \prod_{i=1}^{N} f_{k_{i}}(\vect{x}_{i-1},\vect{\theta},\tau_{i-1})\exp\left(- \tau_{i} \sum_{j'=1}^{M}f_{j'}(\vect{x}_{i-1},\vect{\theta},\tau_{i-1}) \right) $$
where N is the number of transitions occurred in the time interval $[0,T]$, $k_{i}\in[1,\dots M]$ is the type of the $i^{th}$ transition and $\tau_{i}, \vect{x}_{i}$ are the time and state at the $i^{th}$ transition respectively. Notice that the likelihood function corresponds to the generative process described by the SSA. By specifying a suitable prior and applying Bayes' theorem, we can obtain the posterior distribution $p(\vect{\theta}| \vect{X})$ which we can use for inference over the unknown parameters $\vect{\theta}$ \cite{Boys2008}.

In many problems of interest however we cannot observe the times and types of all transitions in a given time interval. Rather, we can only observe the state of the system $\vect{X}(t_{i}) = \vect{x}_{i}$ at discrete time points $t_{i}\in[0,T]$. The solution proposed in \cite{Boys2008} is to treat the trajectories, as well as the number, times and types of transitions, between observed time points as latent variables. This leads to a data augmentation framework \cite{Tanner1987} where a Markov Chain is constructed to sample from the joint posterior of the parameters and the latent variables. At each MCMC iteration the complete trajectory of the MJP process has to be simulated conditional on the observed data and the parameters which for some systems can be computationally demanding. Furthermore, due to the high dimensional nature of the simulated trajectory and the strong dependence on the system parameters and observed data the MCMC algorithm has very poor convergence and mixing properties requiring many samples from the posterior in order to obtain sufficiently accurate Monte Carlo estimates. Finally, a further complication that arises is that the number of transitions between two observed time points is also unknown and has to be sampled using a reversible-jumps type algorithm \cite{Green1995}. For more details see \cite{Boys2008}. The resulting  algorithm therefore is computationally demanding thus limiting its applicability on small and relatively simple MJPs. A more efficient version of the algorithm is also suggested in \cite{Boys2008} where instead of simulating the trajectories between observations using the exact MJP an approximate proposal distribution is employed to sample trajectories which are accepted or rejected using the Metropolis-Hastings ratio.

\section{Diffusion and Linear Noise Approximations}
\label{sec:Approximations}
An alternative to working directly with the master equation and the original MJP is to consider approximations which provide for efficient simulation and possibly an easy to evaluate likelihood function for discretely observed data. Although the resulting posterior will also be approximate in nature, it can be sufficient for inferential purposes given that the system under consideration is near its thermodynamic limit. Here we describe the diffusion approximation and from that how we can arrive at the LNA. Our presentation is rather informal and follows \cite{Wallace2012} and \cite{Gillespie2007}. For a more formal derivation the reader should refer to \cite{vanKampen1992} and \cite{Gillespie2000}. The requirement for these approximations to be consistent is the existence of a proportionality constant $\Omega$ which governs the size of the fluctuations such that for large $\Omega$ the jumps will be relatively small and as both $\Omega$ and $\vect{x}$ tend to infinity approaching the system's thermodynamic limit then, 
\begin{equation}
f_{j}\left(\vect{x},\vect{\theta},t\right) \rightarrow \Omega\tilde{f}_{j}(\vect{z},\vect{\theta},t), \label{eq:varchange}
\end{equation}
where $\vect{z} = \vect{x}/\Omega$ and $\tilde{f}_{j}(\cdot)$ are independent of $\Omega$. For many physical processes where the fluctuations are due to the discrete nature of matter there is a natural $\Omega$ parameter with such properties. Examples of such parameters can be the system size in chemical kinetics, the capacity of a condenser in electric circuits or the mass of a particle \cite{vanKampen1992}.

\subsection{Diffusion approximation}
In order to obtain a Langevin equation which closely matches the dynamics of the MJP it is assumed that there is an infinitesimal time interval $dt$ which satisfies the following conditions
\begin{eqnarray}
f_{j}(\vect{x}_{t'},\vect{\theta},t') \approx f_{j}(\vect{x}_{t},\vect{\theta},t),& \quad &\forall t' \in [t,t+dt), \forall j\in[1,M] \label{eq:tauCondition1}\\
f_{j}(\vect{x}_{t},\vect{\theta},t)dt \gg 1 &\quad &\forall j\in[1,M]. \label{eq:tauCondition2}
\end{eqnarray}
The first condition constrains $dt$ to be small enough such that the transition rate functions remain approximately constant. This implies that the number of transitions of type $j$ is distributed as a Poisson random variable with mean $f_{j}(\vect{x}_{t},\vect{\theta},t)dt$ and is independent from other transitions of type $j'\neq j$. The second condition constrains $dt$ to be large enough such that the number of transitions for each state is significantly larger than 1, which further implies that the Poisson distribution can be accurately approximated by a Gaussian distribution. It can be shown \cite{Gillespie2009} that we can choose $dt$ and $\Omega$ such that both conditions can be satisfied and this generally occurs when the system approaches its thermodynamic limit.

Given such a timescale, the state of the system at time $t+dt$ can be computed by 
\begin{equation}
\vect{x}_{t+dt} = \vect{x}_{t} + \sum_{j=1}^{M}\Norm[f_{j}(\vect{x}_{t},\vect{\theta},t)dt, f_{j}(\vect{x}_{t},\vect{\theta},t)dt]\vect{s}_{j}
\label{eq:tauleaping}
\end{equation}
where $\mathcal{N}[\mu,\sigma^{2}]$ denotes a Gaussian random variate with mean $\mu$ and variance $\sigma^{2}$. From Equation~(\ref{eq:tauleaping}) we can directly obtain a Langevin equation of the form
\begin{equation}
d\vect{x}_{t} = \vect{S}\vect{f}(\vect{x}_{t},\vect{\theta},t)dt+\vect{S}\sqrt{\mbox{diag}[\vect{f}(\vect{x}_{t},\vect{\theta},t)]}d\vect{B}_{t}
\label{eq:Langevin}
\end{equation}
where we used $\vect{S}$ to denote the matrix whose columns are the state change vectors $\vect{s}_{j}$, $\vect{f}(\cdot)$ to denote the vector whose elements are the transition rates $f_{j}(\cdot)$, $\mbox{diag}(\vect{v})$ a function that returns a diagonal matrix with elements taken from the vector $\vect{v}$ and $d\vect{B}_{t}$ an $M$ dimensional Wiener process. Notice that the dimension of $\vect{x}_{t}$ differs from that of $d\vect{B}_{t}$.

Due to the nonlinear state dependent drift and diffusion coefficients in Equation~(\ref{eq:Langevin}) the transition density of the stochastic process is also intractable. Therefore a data augmentation approach similar to the one in \cite{Boys2008} has to be followed. However, there is no longer the need to sample the number, times and types of state transitions as the MJP is approximated with a continuous process. Moreover, the latent variables corresponding to unobserved states can now be efficiently simulated by an Euler-Maruyama scheme which is computationally more efficient than the SSA. This approach has been followed by \cite{Golightly2011} and \cite{Roberts2001} for inference over the unknown parameters $\vect{\theta}$ while in \cite{Heron2007} a similar methodology has been applied on a real data from an auto-regulatory gene expression network.

\subsection{Linear noise approximation}
Substituting equation~(\ref{eq:varchange}) in the Langevin equation~(\ref{eq:Langevin}) and dividing by $\Omega$ we get
\begin{equation}
d\vect{z}_{t} = \vect{S}\tilde{\vect{f}}(\vect{z}_{t},\vect{\theta},t)dt+\frac{1}{\sqrt{\Omega}}\vect{S}\sqrt{\mbox{diag}[\tilde{\vect{f}}(\vect{z}_{t},\vect{\theta},t)]}d\vect{B}_{t}
\label{eq:SDEconsentrationsForm}
\end{equation}
from which we can see that the fluctuations are of the order of $1/\sqrt{\Omega}$ and in the thermodynamic limit (\ref{eq:SDEconsentrationsForm}) reduces to the Macroscopic Rate Equation (MRE)
\begin{equation*}
\lim_{\Omega\rightarrow\infty}d\vect{z}_{t} = \vect{S}\tilde{\vect{f}}(\vect{z}_{t},\vect{\theta},t)dt.
\end{equation*}
To obtain the Linear Noise Approximation (LNA) we make the assumption that for sufficiently large $\Omega$ a solution to (\ref{eq:SDEconsentrationsForm}) will differ from the MRE by a stochastic term of order $1/\sqrt{\Omega}$. That is
\begin{equation}
\vect{z}_{t} = \vect{\phi}_{t} + \frac{1}{\sqrt{\Omega}}\vect{\xi}_{t}
\label{eq:ansatz}
\end{equation}
where $\vect{\phi}_{t}$ are deterministic or sure variables satisfying the MRE and $\vect{\xi}_{t}$ are stochastic variables. Rewriting the transition rate functions using (\ref{eq:ansatz}) and Taylor expand around $\vect{\phi}$ we get
\begin{equation}
\tilde{f}_{j}(\vect{z},\vect{\theta},t) = \tilde{f}_{j}\left(\vect{\phi}+\frac{1}{\sqrt{\Omega}}\vect{\xi}\right) = \tilde{f}_{j}(\vect{\phi},\vect{\theta},t)+\frac{1}{\sqrt{\Omega}}\sum_{d=1}^{D}\frac{\partial \tilde{f}_{j}(\vect{\phi},\vect{\theta},t)}{\partial \phi_{i} }\xi_{i} + O(\Omega^{-1}).
\label{eq:RateFTaylor}
\end{equation}
We can now substitute (\ref{eq:ansatz}) and (\ref{eq:RateFTaylor}) back into (\ref{eq:SDEconsentrationsForm}) and collect terms of $O(1)$ to get the expression for the differential of $\vect{\phi}$ which is nothing other than the MRE
\begin{equation}
d\vect{\phi}_{t} = \vect{S}\tilde{\vect{f}}(\vect{\phi}_{t},\vect{\theta},t)dt.
\label{eq:MRE}
\end{equation}
Finally, collecting remaining terms and neglecting terms of $O(1/\sqrt{\Omega})$ and higher we get the differential of $\vect{\xi}$ as
\begin{equation}
d\vect{\xi}_{t} = \vect{S}\vect{J}_{\tilde{f}}(\vect{\phi}_{t},\vect{\theta},t)\vect{\xi}_{t}dt + \vect{S}\sqrt{\mbox{diag}[\tilde{\vect{f}}(\vect{\phi}_{t},\vect{\theta},t)]}d\vect{B}_{t}
\label{eq:LinearSDE}
\end{equation}
where we used $\vect{J}_{\tilde{f}}(\cdot)$ to denote the Jacobian of the transition rates $\tilde{\vect{f}}(\cdot)$. Equation~(\ref{eq:LinearSDE}) characterises the fluctuations around the deterministic state $\vect{\phi}$ and its validity depends on the size of $\Omega$. As $\Omega$ increases the magnitude of the individual jumps $\vect{s}_{j}$ becomes negligible relative to the distance in $\vect{\phi}$ over which the non-linearity of $\tilde{f}_{j}(\cdot)$ becomes noticeable. A measure of the sufficiency of  LNA is the coefficient of variation, i.e. the ratio of the standard deviation to the mean. For a more thorough discussion on the validity of LNA the reader is referred to \cite{Ferm2008} and the supplementary material of \cite{Komorowski2009}. 

\subsection{Solution of the LNA and the approximate likelihood function}
LNA provides a convenient expression for the approximate likelihood since the MRE~(\ref{eq:MRE}) can be easily solved numerically and its computational cost is polynomial in $D$. Moreover, equation~(\ref{eq:LinearSDE}) is a system of linear stochastic differential equations which has an explicit solution of the form
\begin{equation}
\vect{\xi}_{t} = \vect{\Phi}(t_{0},t)\left(\xi_{0} +\int_{t_{0}}^{t}\vect{\Phi}(s,t)^{-1}\vect{S}\sqrt{\mbox{diag}[\tilde{\vect{f}}(\vect{\phi}_{s},\vect{\theta},s)]}d\vect{B}_{s} \right)
\label{eq:XiSolution}
\end{equation}
where the integral is in the It\^o sense and $\vect{\Phi}(t_{0},t)$ is the solution of 
\begin{equation}
d\vect{\Phi}(t_{0},s) =  \vect{S}\vect{J}_{\tilde{f}}(\vect{\phi}_{t},\vect{\theta},t)\vect{\Phi}(t_{0},s)ds,\quad \vect{\Phi}(t_{0},t_{0}) = \vect{I}.
\label{eq:PhiMatLNA}
\end{equation}
Since the It\^o integral of a deterministic function is a Gaussian random variable \cite{Oksendal1992}, equation~(\ref{eq:XiSolution}) implies that $\vect{\xi}_{t}$ has a multivariate normal distribution. To simplify further the analysis assume that the initial condition for $\vect{z}_{t}$ has a multivariate normal distribution such that $\vect{z}_{t_{0}}\sim\Norm(\vect{\phi}_{t_{0}},\vect{V}_{t_{0}})$. For the rest of the paper we will assume that $\vect{\phi}_{t_{0}}$ and $\vect{V}_{t_{0}}$ are known. In cases where the initial conditions are unknown they can be treated as additional parameters. Equations~(\ref{eq:ansatz}, \ref{eq:MRE}, \ref{eq:LinearSDE}, \ref{eq:XiSolution}) and the specification of initial conditions further imply that 
\begin{equation}
\vect{z}_{t} \sim \Norm(\vect{\phi}_{t},\Omega^{-1}\vect{V}_{t})
\label{eq:LNASolution}
\end{equation}
where $\vect{\phi}_{t}$ are solutions of the MRE and $\vect{V}_{t}$ are solutions of 
$$d\vect{V}_{t} = \vect{S}\vect{J}_{\tilde{f}}(\vect{\phi}_{t},\vect{\theta},t)\vect{V}_{t} + \vect{V}_{t}\vect{J}^{T}_{\tilde{f}}(\vect{\phi}_{t},\vect{\theta},t)\vect{S}^{T}+\vect{S}\mbox{diag}[\tilde{\vect{f}}(\vect{\phi}_{t},\vect{\theta},t)]\vect{S}^{T}.$$
Finally, multiplying (\ref{eq:LNASolution}) by $\Omega$ we get  
$$\vect{x}_{t} \sim \Norm(\Omega\vect{\phi}_{t},\Omega\vect{V}_{t}).$$

Assume that we have observations from the stochastic process $\vect{X}(t)$ at discrete time points $t_{i} \in \{t_{1},\dots, t_{N}\}$. Moreover, assume that each observation $\vect{x}_{t}$ is obtained by a independent realisation of $\vect{X}(t)$. For example to obtain an observation at $t_{1} = 10$ the SSA is used to simulate a trajectory from $t_{0}$ to $t_{1}$ and the state of the system at $t_{1}$ is kept. For $t_{2} = 20$ the SSA is again used to simulate a new trajectory from $t_{0}$ to $t_{2}$ keeping only the state of the system at $t_{2}$ and the process continues until all necessary observations are gathered. This kind of data are very frequently encountered in biology where in order to obtain a single measurement the sample has to be ``sacrificed''. This is common in data obtained using Polymerase Chain Reaction reporter assays \cite{Nolan2006} for example. See also \cite{Komorowski2009} for an example of an inference problem with such data. Due to the independence between different observations and the Markov property the likelihood is simply
\begin{equation}
p(\vect{X}|\vect{\theta}) = \prod_{i=1}^{N}\Norm(\vect{x}_{t_{i}}|\Omega\vect{\phi}_{t_{i}},\Omega\vect{V}_{t_{i}}).
\label{eq:LNALikelihood}
\end{equation}

In this paper we only consider observations of this kind. However the methodology is readily applicable when observations from a single realisation of $\vect{X}(t)$ are available. In this case the likelihood also has a simple form 
$$p(\vect{X}|\vect{\theta}) = \Norm[\vect{X}|\Omega\vect{\mu}(\vect{\theta}),\Omega\vect{\Sigma}(\vect{\theta})]$$
where $\vect{X}= (\vect{x}_{t_{1}},\dots,\vect{x}_{t_{N}})^{T}$ is an $ND$ vector with all the observations, $\vect{\mu}(\vect{\theta})=(\vect{\phi}_{t_{1}},\dots,\vect{\phi}_{N})^{T}$, is also a $ND$ vector with solutions of the MRE and $\vect{\Sigma}(\vect{\theta})$ is a $ND\times ND$ block matrix $\vect{\Sigma}(\vect{\theta}) = \{ \vect{\Sigma}(\vect{\theta})^{i,j}: i,j\in[1,\dots,N] \}$ such that 
\begin{equation}
\vect{\Sigma}(\vect{\theta})^{i,j} = \left\{ \begin{array}{cc} \vect{V}_{t_{i}}, & i=j \\ \vect{V}_{t_{i}}\vect{\Phi}(t_{i},t_{j})^{T}, & i\neq j \end{array}\right.
\label{eq:varLNA}
\end{equation} 
This stems from the fact that due to the Markov property and equation~(\ref{eq:LNASolution}) each $\vect{x}_{t_{i}}$ can be written as a sum of multivariate normal random variables and therefore $\vect{X}$ is also a multivariate normal random variable. For more details refer to the supplementary material of \cite{Komorowski2009} and \cite{Komorowski2011}. The only additional complication which arises for time-series data is that the off-diagonal components of the LNA variance in equation (\ref{eq:varLNA}) need to be estimated by numerically solving the system of ODEs in equation (\ref{eq:PhiMatLNA}). Notice that despite the fact that the variance matrix is full we can still exploit the Markov property and write the likelihood as a product of the conditional likelihoods and therefore avoid the cost of inverting the $ND\times ND$ variance matrix.

\section{Markov Chain Monte Carlo Methods}
\label{sec:MCMC}
In this section we give a brief overview of the MCMC algorithms that we consider in this work. Some familiarity with the concepts of MCMC is required by the reader since an introduction to the subject is out of the scope of this paper. 

\subsection{Metropolis-Hastings}
For a random vector $\boldsymbol{\theta} \in \mathbb{R}^D$ with density $p(\boldsymbol{\theta})$ the Metropolis-Hastings algorithm employs a proposal mechanism $q(\boldsymbol{\theta}^{*}|\boldsymbol{\theta}^{t-1})$ and proposed moves are accepted with probability $$\min \left\{1,p(\boldsymbol{\theta}^{*}) q(\boldsymbol{\theta}^{t-1}|\boldsymbol{\theta}^{*})/p(\boldsymbol{\theta}^{t-1})q(\boldsymbol{\theta}^{*}|\boldsymbol{\theta}^{t-1})\right\}$$. In the context of Bayesian inference the target density $p(\boldsymbol{\theta})$ corresponds to the posterior distribution of the model parameters. Tuning the Metropolis-Hastings algorithm involves selecting the right proposal mechanism. A common choice is to use a random walk Gaussian proposal of the form $q(\boldsymbol{\theta}^{*}|\boldsymbol{\theta}^{t-1}) = \mathcal{N}(\boldsymbol{\theta}^{*}|\boldsymbol{\theta}^{t-1},\boldsymbol{\Sigma})$, where $\mathcal{N}(\cdot|\boldsymbol{\mu},\boldsymbol{\Sigma})$ denotes the multivariate normal density with mean $\boldsymbol{\mu}$ and covariance matrix $\boldsymbol{\Sigma}$.

Selecting the covariance matrix however, is far from trivial in most cases since knowledge about the target density is required. Therefore a more simplified proposal mechanism is often considered where the covariance matrix is replaced with a diagonal matrix such as $\boldsymbol{\Sigma}=\epsilon\boldsymbol{I}$ where the value of the scale parameter $\epsilon$ has to be tuned in order to achieve fast convergence and good mixing. Small values of $\epsilon$ imply small transitions and result in high acceptance rates while the mixing of the Markov Chain is poor. Large values on the other hand, allow for large transitions but they result in most of the samples being rejected. 

Tuning the scale parameter becomes even more difficult in problems where the standard deviations of the marginal posteriors differ substantially, since different scales are required for each dimension, and this is exacerbated when correlations between different variables exist. Adaptive schemes for the Metropolis-Hastings algorithm have also been proposed \cite{Haario2005} though they should be applied with care \cite{Andrieu2008}. Parameters such as reaction rate constants often differ orders of magnitude, thus a scaled diagonal covariance matrix will be a bad choice for such problems. In the numerical simulations in the next section we used a Metropolis within Gibbs scheme where each parameter is updated conditional on all others using a univariate normal density with a parameter-specific scale parameter. This allows us to tune the scale for each proposal independently and achieve better mixing.

\subsection{Manifold Metropolis Adjusted Langevin Algorithm}
Denoting the log of the target density as $\mathcal{L}(\boldsymbol{\theta}) = \log p(\boldsymbol{\theta})$, the manifold MALA (MMALA) method, \cite{Girolami2011}, defines a Langevin diffusion with stationary distribution $p(\boldsymbol{\theta})$ on the Riemann manifold of density functions with metric tensor $\boldsymbol{G}(\boldsymbol{\theta})$. By employing a first order Euler integrator to solve the diffusion a proposal mechanism with density 
$q(\boldsymbol{\theta}^*|\boldsymbol{\theta}^{t-1}) = \mathcal{N}(\boldsymbol{\theta}^*| \boldsymbol{\mu}(\boldsymbol{\theta}^{t-1},\epsilon),\epsilon^2\boldsymbol{G}^{-1}(\boldsymbol{\theta}^{t-1}))$
is obtained, where $\epsilon$ is the integration step size, a parameter which needs to be tuned, and the $d$th component of the mean function $\boldsymbol{\mu}(\boldsymbol{\theta},\epsilon)_d$ is
\begin{eqnarray}
\boldsymbol{\mu}(\boldsymbol{\theta},\epsilon)_d &  = & \boldsymbol{\theta}_d + \frac{\epsilon^2}{2}\left(\boldsymbol{G}^{-1}(\boldsymbol{\theta})\nabla_{\boldsymbol{\theta}}\mathcal{L}(\boldsymbol{\theta})\right)_d - \epsilon^2 \sum_{i=1}^D\sum_{j=1}^D\boldsymbol{G}(\boldsymbol{\theta})_{i,j}^{-1}\Gamma_{i,j}^d  \label{eq:meanmMALA}
\end{eqnarray}
where $\Gamma_{i,j}^d$ are the Christoffel symbols of the metric in local coordinates \cite{Kuhnel2005}.

Similarly to MALA \cite{Roberts2003}, due to the discretisation error introduced by the first order approximation, convergence to the stationary distribution is not guaranteed anymore and thus the Metropolis-Hastings ratio is employed to correct this bias. The MMALA algorithm can be simply stated as in Algorithm \ref{alg:mala} and more details can be found in \cite{Girolami2011}. 
\algsetup{indent=2em} 
\begin{algorithm}[h!] \caption{MMALA}\label{alg:mala} 
\begin{algorithmic}[1]
\medskip
\STATE Inititialise $\boldsymbol{\theta}^0$
\FOR{$t=1$ to $T$}
\STATE $\boldsymbol{\theta}^{*} \sim  \mathcal{N}(\boldsymbol{\theta}| \boldsymbol{\mu}(\boldsymbol{\theta}^{t-1},\epsilon),\epsilon^2\boldsymbol{G}^{-1}(\boldsymbol{\theta}^{t-1}))$
\STATE $r = \min \left\{1,p(\boldsymbol{\theta}^{*}) q(\boldsymbol{\theta}^{t-1}|\boldsymbol{\theta}^{*})/p(\boldsymbol{\theta}^{t-1})q(\boldsymbol{\theta}^{*}|\boldsymbol{\theta}^{t-1})\right\}$
\STATE $u \sim \mathcal{U}_{[0,1]}$
\IF {$r>u$} 
\STATE $\boldsymbol{\theta}^t = \boldsymbol{\theta}^*$
\ELSE
\STATE $\boldsymbol{\theta}^t = \boldsymbol{\theta}^{t-1}$
\ENDIF
\ENDFOR
\end{algorithmic}
\end{algorithm}

We can interpret the proposal mechanism of MMALA as a local Gaussian approximation to the target density similar to the adaptive Metropolis-Hastings of \cite{Haario1998}. In contrast to \cite{Haario1998}, the effective covariance matrix in MMALA is the inverse of the metric tensor evaluated at the current position and no samples from the chain are required in order to estimate it, therefore avoiding the difficulties of adaptive MCMC discussed in \cite{Andrieu2008}.
Furthermore a simplified version of the MMALA algorithm (SMMALA) can also be derived by assuming a manifold with constant curvature, thus cancelling the last term in Equation (\ref{eq:meanmMALA}) which depends on the Christoffel symbols. Finally, the MMALA algorithm can be seen as a generalisation of the original MALA \cite{Roberts2003} since, if the metric tensor $\boldsymbol{G}(\boldsymbol{\theta})$ is equal to the identity matrix corresponding to an Euclidean manifold, then the original algorithm is recovered.

\subsection{Manifold Hamiltonian Monte Carlo}
The Riemann manifold Hamiltonian Monte Carlo (RMHMC) method defines a Hamiltonian on the Riemann manifold of probability density functions by introducing the auxiliary variables $\boldsymbol{p}\sim \mathcal{N}(\boldsymbol{0},\boldsymbol{G}(\boldsymbol{\theta}))$, which are interpreted as the momentum at a particular position $\boldsymbol{\theta}$ and by considering the negative log of the target density as a potential function. More formally, the Hamiltonian defined on the Riemann manifold is:
\begin{equation}
H(\boldsymbol{\theta},\boldsymbol{p}) = -\mathcal{L}(\boldsymbol{\theta}) +\frac{1}{2}\log\left(2\pi|\boldsymbol{G}(\boldsymbol{\theta})|\right) + \frac{1}{2}\boldsymbol{p}^T\boldsymbol{G}(\boldsymbol{\theta})^{-1}\boldsymbol{p}
\end{equation}
where the terms $-\mathcal{L}(\boldsymbol{\theta}) +\frac{1}{2}\log\left(2\pi|\boldsymbol{G}(\boldsymbol{\theta})|\right) $ and $\frac{1}{2}\boldsymbol{p}^T\boldsymbol{G}(\boldsymbol{\theta})^{-1}\boldsymbol{p}$ are the potential energy and kinetic energy terms, respectively.
Simulating the Hamiltonian requires a time-reversible and volume preserving numerical integrator. For this purpose the Generalised Leapfrog algorithm can be employed and provides a deterministic proposal mechanism for simulating from the conditional distribution, i.e. $\boldsymbol{\theta}^*|\boldsymbol{p} \sim p(\boldsymbol{\theta}^*|\boldsymbol{p})$. More details about the Generalised Leapfrog integrator can be found in \cite{Girolami2011}. To simulate a path across the manifold, the Leapfrog integrator is iterated $L$ times which along with the integration step size $\epsilon$ are parameters requiring tuning. Again, due to the integration errors on simulating the Hamiltonian, in order to ensure convergence to the stationary distribution the Metropolis-Hastings ratio is applied. Moreover, following the suggestion in \cite{Neal93} the number of Leapfrog iterations $L$ is randomised in order to improve mixing. The RMHMC algorithm is given in Algorithm \ref{alg:rmhmc}. 
\algsetup{indent=2em} 
\begin{algorithm}[h!] \caption{RMHMC}\label{alg:rmhmc} 
\begin{algorithmic}[1]
\medskip
\STATE Inititialise $\boldsymbol{\theta}^0$
\FOR{$t=1$ to $T$}
\STATE $\boldsymbol{p}^{0}_* \sim \mathcal{N}(\boldsymbol{p}| \boldsymbol{0},\boldsymbol{G}(\boldsymbol{\theta}^{t-1}))$
\STATE $\boldsymbol{\theta}_*^0= \boldsymbol{\theta}^{t-1}$
\STATE $e \sim \mathcal{U}_{[0,1]}$
\STATE $N = \ensuremath{\mbox{ceil}}(\epsilon L)$
\\\COMMENT{Simulate the Hamiltonian using a generalised Leapfrog integrator for N steps}
\FOR{$n=0$ to $N$}
\STATE solve $\boldsymbol{p}^{n+\frac{1}{2}}_* =  \boldsymbol{p}^{n}_* - \frac{\epsilon}{2}\nabla_{\boldsymbol{\theta}}H\left(\boldsymbol{\theta}^n_*,\boldsymbol{p}^{n+\frac{1}{2}}_*\right)$
\STATE solve $\boldsymbol{\theta}^{n+1}_* = \boldsymbol{\theta}^n_* +\frac{\epsilon}{2}\left[ \nabla_{\boldsymbol{p}}H\left(\boldsymbol{\theta}^n_*,\boldsymbol{p}^{n+\frac{1}{2}}_*\right) + \nabla_{\boldsymbol{p}}H\left(\boldsymbol{\theta}^{n+1}_*,\boldsymbol{p}^{n+\frac{1}{2}}_*\right) \right] $
\STATE  $\boldsymbol{p}^{n+1}_* = \boldsymbol{p}^{n+\frac{1}{2}}_* - \frac{\epsilon}{2}\nabla_{\boldsymbol{\theta}}H\left(\boldsymbol{\theta}^{n+1}_*,\boldsymbol{p}^{n+\frac{1}{2}}_*\right) $
\ENDFOR
\STATE $\left( \boldsymbol{\theta}^*,\boldsymbol{p}^* \right) = \left( \boldsymbol{\theta}^{N+1}_*,\boldsymbol{p}^{N+1}_* \right)$
\\\COMMENT{Metropolis-Hastings ratio}
\STATE $r = \min\left\{1, \exp\left(-H(\boldsymbol{\theta}^{*},\boldsymbol{p}^{*}) +H(\boldsymbol{\theta}^{t-1},\boldsymbol{p}^{t-1})\right) \right\}$
\STATE $u \sim \mathcal{U}_{[0,1]}$
\IF {$r>u$} 
\STATE$ \boldsymbol{\theta}^t =  \boldsymbol{\theta}^*$
\ELSE
\STATE $\boldsymbol{\theta}^t = \boldsymbol{\theta}^{t-1}$
\ENDIF
\ENDFOR
\end{algorithmic}
\end{algorithm}

\noindent Similar to the MMALA algorithm, when the metric tensor $\boldsymbol{G}(\boldsymbol{\theta})$ is equal to the identity matrix corresponding to an Euclidean manifold, then RMHMC is equivalent to the HMC algorithm of \cite{Duane1987}.

\section{Implementation details}
\subsection{Gradient and metric tensor for the LNA}
For the manifold MCMC algorithms discussed in this section we will need the gradient of the log likelihood as well as a metric tensor for the LNA. For density functions the natural metric tensor is the expected Fisher Information, $\boldsymbol{I}(\boldsymbol{\theta})$, \cite{Amari2000} and  for a multivariate normal with mean $\vect{\mu}(\vect{\theta})$ and covariance matrix $\vect{\Sigma}(\vect{\theta})$ its general form  is
$$I(\vect{\theta})_{i,j} = \frac{\partial\vect{\mu}(\vect{\theta})}{\partial\theta_{i}}\vect{\Sigma}^{-1}(\vect{\theta}) \frac{\partial\vect{\mu}(\vect{\theta})}{\partial\theta_{j}} +\frac{1}{2}\mbox{Tr}\left(\vect{\Sigma}^{-1}(\vect{\theta})\frac{\partial \vect{\Sigma}(\vect{\theta})}{\partial \theta_{i}}\vect{\Sigma}^{-1}(\vect{\theta})\frac{\partial \vect{\Sigma}(\vect{\theta})}{\partial \theta_{j}}\right).$$
For the likelihood in equation~(\ref{eq:LNALikelihood}) the Fisher Information is then a sum of $N$ matrices $\vect{I}(\vect{\theta},t)$, one evaluated at each time point. Similarly the general form of the partial derivatives for the log of a multivariate normal is 
$$ \frac{\partial\ln\Norm[\vect{x}|\vect{\mu}(\vect{\theta}),\vect{\Sigma}(\vect{\theta}) ]}{\partial \theta_{i}} = \frac{1}{2}\mbox{Tr}\left[ (\vect{c}\vect{c}^{T} - \vect{\Sigma}^{-1}(\vect{\theta}) )\frac{\partial \vect{\Sigma}(\vect{\theta})}{\partial \theta_{i}}\right] + \vect{c}^{T}\frac{\partial\vect{\mu}(\vect{\theta})}{\partial\theta_{i}} $$
where $\vect{c} = \vect{\Sigma}^{-1}(\vect{\theta})[\vect{x} - \vect{\mu}(\vect{\theta})]$.

Moreover, during the leap-frog integration for the RMHMC and for the mean function of MMALA the partial derivatives of the Fisher Information are needed. Their general form is 
\begin{eqnarray*}
\frac{\partial I(\vect{\theta})_{i,j}}{\partial\theta_{k}} &=& \frac{\partial^{2}\vect{\mu}(\vect{\theta})^{T} }{\partial \theta_{i}\partial \theta_{k} }\vect{a}_{j}+ \vect{a}_{i}^{T}\frac{\partial^{2}\vect{\mu}(\vect{\theta}) }{\partial \theta_{j}\partial \theta_{k} } -\vect{a}_{i}^{T}\frac{\partial \vect{\Sigma}(\vect{\theta})}{\partial \theta_{k}}\vect{a}_{j}\\
&-& \frac{1}{2}\mbox{Tr}\left[\vect{A}_{k}(\vect{A}_{i}\vect{A}_{j} +\vect{A}_{j}\vect{A}_{i})\right]\\
 &+& \frac{1}{2}\mbox{Tr}\left[\vect{\Sigma}^{-1}(\vect{\theta})\left(\frac{\partial \vect{\Sigma}(\vect{\theta})}{\partial \theta_{i}\partial\theta_{k}}\vect{A}_{j} + \frac{\partial \vect{\Sigma}(\vect{\theta})}{\partial \theta_{j}\partial\theta_{k}}\vect{A}_{i} \right) \right]
\end{eqnarray*}
where $\vect{a}_{i} = \vect{\Sigma}^{-1}\frac{\partial\vect{\mu}(\vect{\theta})}{\partial \theta_{i}}$ and $\vect{A}_{i} = \vect{\Sigma}^{-1}\frac{\partial\vect{\Sigma}(\vect{\theta})}{\partial \theta_{i}}$. 

The above quantities require first and second order sensitivities for the $\vect{\phi}$ and $\vect{V}$ which we obtain by augmenting the ODE systems with the additional sensitivity equations. For an ODE system of $n_{y}$ equations with form $\dot{\vect{y}}=\vect{F}(\vect{y},t,\vect{\theta}), \quad \vect{y}(t_{0}) = \vect{y}_{0}(\vect{\theta})$ and $n_{\theta}$ parameters $\vect{\theta}$, the first and second order forward sensitivity equations are given by (\ref{eg:firstSens}) and (\ref{eg:secSens}) respectively.
\begin{equation}
\frac{\partial \dot{\vect{y}}}{\partial\vect{\theta}} = \vect{F}_{\vect{y}}\frac{\partial \vect{y} }{\partial \vect{\theta}} + \vect{F}_{\vect{\theta}}, \quad \frac{\partial \vect{y}(t_{0})}{\partial \vect{\theta}}=\frac{\partial \vect{y}_{0}}{\partial \vect{\theta}}
\label{eg:firstSens}
\end{equation}
\begin{eqnarray}
\frac{\partial^{2} \dot{\vect{y}}}{\partial\vect{\theta}\partial\vect{\theta}^{T}} &=& [\vect{F}_{\vect{y}}\otimes\vect{I}_{n_{\theta}}]\frac{\partial^{2} \vect{y}}{\partial\vect{\theta}\partial\vect{\theta}^{T}} + \left[\vect{I}_{n_{y}}\otimes\frac{\partial \vect{y}^{T}}{\partial \vect{\theta}} \right]\left[\vect{F}_{\vect{y},\vect{y}}\frac{\partial \vect{y} }{\partial \vect{\theta}} + \vect{F}_{\vect{y},\vect{\theta}}\right] \nonumber \\ &+& \left[\vect{F}_{\vect{\theta},\vect{y}}\frac{\partial \vect{y} }{\partial \vect{\theta}} + \vect{F}_{\vect{\theta},\vect{\theta}}\right], \nonumber  \\
& & \frac{\partial^{2} \vect{y}(t_{0})}{\partial\vect{\theta}^{2}} = \frac{\partial^{2} \vect{y}_{0}}{\partial\vect{\theta}^{2}} 
\label{eg:secSens}
\end{eqnarray}

We use $\vect{F}_{\vect{\theta}}$ to denote the $n_{y}\times n_{\theta}$ matrix where its $j^{th}$ column is the partial derivatives of $\vect{F}$ with respect to $\theta_{j}$. $\vect{F}_{\vect{\theta},\vect{y}}$ denotes the derivative of $\vect{F}_{\vect{\theta}}$ with respect to $\vect{y}$ and is an $n_{\theta}\cdot n_{y}\times n_{y}$ matrix where its $j^{th}$ column is the partial derivatives of $\mbox{vec}(\vect{F}^{T}_{\vect{\theta}})$ with respect to $y_{j}$. $I_{n_{y}}$ denotes the $n_{y}\times n_{y}$ Identity matrix, $\otimes$ the Kronecker product and $\mbox{vec}(\vect{A})$ an operator that creates a column vector by stacking the columns of matrix $\vect{A
}$.

\subsection{Re-parameterisation}
In many problems the parameters $\vect{\theta}$ can be constrained in certain parts of $\mathbb{R}^{n_{\theta}}$ where $n_{\theta}$ is the number of parameters. In models of chemical kinetics for example, rate parameters must be positive and can differ by orders of magnitude. For the MCMC algorithms described in the previous section we will need a re-parameterisation in order to allow the algorithms to operate on an unbounded and unconstrained parameter space.

For the numerical simulations in section~\ref{sec:experiments} we use a $\log_{10}$ re-parameterisation by introducing the variables $\check{\theta}_{p} = \log_{10}(\theta_{p})$, $p\in[1,\dots,n_{\theta}]$. To ensure that we sample from the correct posterior the joint density is scaled by the determinant of the Jacobian such that $p(\vect{X}|\check{\vect{\theta}})p(\check{\vect{\theta}})|\vect{J}(\check{\vect{\theta}})|$ where 
$\vect{J}(\check{\vect{\theta}})$ is a $n_{\theta}\times n_{\theta}$ diagonal matrix with elements $\vect{J}(\check{\vect{\theta}})_{p,p} = 10^{\check{\theta}_{p}}\log(10)$.

The gradient and  Fisher information along with its partial derivatives follow from the chain rule as
\begin{eqnarray*}
\nabla_{\check{\vect{\theta}}}\mathcal{L}(\check{\vect{\theta}}) &=& \nabla_{\vect{\theta}}\mathcal{L}(\vect{\theta})\vect{J}(\check{\vect{\theta}})\\
\vect{I}(\check{\vect{\theta}})& =&  \vect{J}(\check{\vect{\theta}})^{T}\vect{I}(\vect{\theta})\vect{J}(\check{\vect{\theta}})\\
\frac{\partial \vect{I}(\check{\vect{\theta}}) }{\partial \check{\theta}_{p}} &=& 2\vect{J}(\check{\vect{\theta}})^{T}\vect{I}(\vect{\theta})\frac{\partial \vect{J}(\check{\vect{\theta}}) }{\partial \check{\theta}_{p}} + \vect{J}(\check{\vect{\theta}})^{T}\frac{\partial\vect{I}(\vect{\theta})}{\partial \theta_{p}}\vect{J}(\check{\vect{\theta}})\frac{\partial \theta_{p}}{\partial \check{\theta_{p}}}
\end{eqnarray*}

\subsection{Choice of priors}
In Bayesian statistics priors provide the means for incorporating existing knowledge for the parameters in question. The choice of a suitable prior distribution can be informed from knowledge about the process being modelled, the experimental design and empirical observations. For example we might want to restrict rate parameters in chemical kinetics from becoming very high since we assume from the experimental design that reactions are slow enough to be able to be observed. In some cases the model itself can also guide the choice of the prior. For example when a model is only defined for a certain range of values of the parameters, a prior restricting the parameters in that range should be used. 

In the numerical simulations of the next section we use independent normal priors for the parameters $\check{\vect{\theta}}$. Due to the re-parameterisation introduced earlier, this corresponds to a log-normal prior with base 10 for the parameters $\vect{\theta}$. This choice allows parameters to differ several orders of magnitude while it ensures they are strictly positive. Moreover, as noted in \cite{Girolami2011} the negative Hessian of the prior is added to the Fisher information in order to form the metric tensor used during MCMC sampling. This has the added benefit of regularising the Fisher information when it is near-singular \cite{Calderhead2011} although we have not observed such problems in the simulations presented here.

\section{Numerical Simulations}\label{sec:experiments}
\subsection{Chemical kinetics}

In this section we consider two examples from chemical kinetics \cite{Wallace2012} and study the effect of the system size parameter on inference using MCMC. The first system consists of three species where an unstable monomer, $S_{1}$, can dimerise to an unstable dimer, $S_{2}$, which is then converted to a stable form, $S_{3}$. The reaction set for this system is
\begin{eqnarray*}
R1 : S_{1} & \xrightarrow{c_{1}} & \emptyset \\
R2: 2S_{1} & \xrightarrow{c_{2}\Omega^{-1}} &  S_{2}\\ 
R3: S_{2} & \xrightarrow{c_{3}} & 2S_{1}\\
R4: S_{2} & \xrightarrow{c_{4}} & S_{3}\\
\end{eqnarray*}
and the state of the system at time $t$ will be denoted by $\vect{X}(t)=[S_{1}(t), S_{2}(t), S_{3}(t)]^{T}$. The propensity functions, or state transition probabilities are $\vect{f}(\vect{X},\vect{\theta}) = [c_{1}S_{1}(t), c_{2}\Omega^{-1}S_{1}(t)(S_{1}(t)-1)/2, c_{3}S_{2}(t), c_{4}S_{3}(t)]^{T}$ and the corresponding state change matrix is 
\begin{equation}
\vect{S} = \left(\begin{array}{cccc} -1 & -2 & 2 & 0 \\0 & 1 & -1 & -1\\0 & 0 & 0 & 1\end{array}\right).
\label{eq:dimmer:S}
\end{equation}

\begin{table}
\centering
\begin{tabular}{@{\extracolsep{0.0pt}}llll}
\multicolumn{4}{c}{ min. ESS vs. $\Omega$} \\
\toprule
$\Omega$ & M.H. & SMMALA & RMHMC \\
\midrule
 1 & 121 (3.6) & 150 (3.9) &  245 (0.06)\\
2 & 226 (6.7) & 2163 (57.2) & 4775 (1.3)\\
5 & 132 (3.9) & 3539 (93.6) & 4618 (1.2)\\
10 & 180 (5.3) & 3397 (89.8) & 5954 (1.6)\\
100 & 214 (6.4) & 3725 (98.5) & 6066 (1.7)\\
\bottomrule
\end{tabular}
\caption{Comparison of minimum Effective Sample Size (ESS) and time normalised min. ESS for different values of the system size parameter $\Omega$ of the decay dimerisation reaction model. Time normalised ESS is given in parenthesis. Results are calculated from 10,000 posterior samples.\label{table:dimer:ESSOmega}}
\end{table}

For our experiments we will assume that initial conditions are known and set them to $S_{1}(t_{0}) = 5\Omega$, $S_{2}(t_{0})=S_{3}(t_{0}) = 0$, $t_{0}=0$. Moreover we will set the reaction rate parameters to $c_{1}=1$, $\hat{c_{2}}=2\Omega^{-1}$, $c_{3}=0.5$ and $c_{4} =0.04$. Notice that we make explicit the relation between the system size and parameter $\hat{c_{2}}$ and we will infer rate $c_{2}$ up to a proportionality constant. For all the experiments we simulate data using the SSA of \cite{Gillespie2007} for the time interval $t \in[0,10]$ and we discretise such that $t_{i}-t_{i-1} = 0.1$. Each observation $\vect{X}(t_{i})$ is obtained independently by simulating a trajectory from $t_{0}$ to $t_{i}$ and keeping only the last state discarding the rest of the trajectory. Moreover for each time point $t_{i}$ we also simulate 10 independent observations. Since each observation is obtained by a different trajectory of the MJP we assume that initial conditions do not have a point mass rather for each trajectory we sample its initial condition from a Poisson with means $S_{1}(0),S_{2}(0),S_{3}(0)$.

We use the synthetic data to perform inference for the rate parameters $\vect{\theta}=(c_{1},\hat{c}_{2},c_{3},c_{4})^{T}$ by drawing samples from the posterior $$p(\vect{\theta}| \vect{X}) \propto p(\vect{\theta})\prod_{i=1}^{N}\prod_{r=1}^{10}\Norm[\vect{X}_{r}(t_{i}) | \Omega\vect{\phi}(t_{i}),\Omega\vect{V}(t_{i})]$$ where $r$ indexes independent observations for the same time point. For all simulations in this paper we assume that the means for the initial conditions are known. Following similar arguments as for the derivation of the LNA in Section \ref{sec:Approximations}, namely that as the system approaches its thermodynamic limit transition densities become Gaussian, the initial conditions for the ODE systems for the mean and variance of the transition densities are $\vect{\phi}(0) = \vect{X}(0)\Omega^{-1}$ and $\vect{V}(0) = \vect{I}$, where $\vect{I}$ is the identity matrix. In a more realistic scenario the initial conditions must be included as additional parameters in $\vect{\theta}$. For all parameters we used an independent log-normal prior with base 10, zero mean and one standard deviation and chains are initialised by drawing a random sample from the prior. For the Metropolis-Hastings sampler we set the initial proposal scale parameters to $\approx1e^{-6}$ and automatically adapt them every 100 samples during the burn-in phase in order to achieve an acceptance rate of $25\% -30\%$ \cite{Roberts1997}. The same adaptation strategy was followed for the simplified MMALA and RMHMC algorithms where the initial step size was also set to $\approx1e^{-6}$ and was tuned in order to achieve acceptance rates in the order of $70-80\%$ \cite{Girolami2011}. Finally, the number of leap-frog steps for RMHMC was fixed to 5. We have found that a burn in period of 10,000 to 20,000 samples was adequate for all algorithms to converge to the stationary distribution. 

\begin{table}
\centering
\begin{tabular}{@{\extracolsep{0.0pt}}lllll}
\multicolumn{5}{c}{Posterior mean and SD. vs. $\Omega$} \\
\toprule
$\Omega$& $c_{1}$ & $\hat{c}_{2}$ & $c_{3}$ & $c_{4}$ \\
\midrule
True & 1 & 2$\Omega^{-1}$ &0.5& 0.04\\
\midrule
1&  0.88 (0.031) & 1.72  (0.253) & 0.39 (0.039)  & 0.003 (0.002) \\
2 &  1.3 (0.041)  & 0.69 (0.066) & 0.35 (0.016) & 0.014 (0.002) \\
5 & 0.93 (0.019) & 0.39 (0.028) & 0.48 (0.025)  & 0.034 (0.002) \\
10 & 1.0 (0.015) & 0.18 (0.008) & 0.47 (0.015) & 0.037 (0.001) \\
100 & 0.99 (0.004) & 0.01 (0.0002) & 0.52 (0.004) & 0.039 (0.0003) \\
\bottomrule
\end{tabular}
\caption{Marginal posterior means and standard deviations calculated from the RMHMC chain for different values of the system size parameter $\Omega$ of the decay-dimerisation reaction model. Notice that $\hat{c}_{2}$ parameter is proportional to $\Omega$. Results are calculated from 10,000 posterior samples.\label{table:dimer:MeanSDRMHMC}}
\end{table}

\begin{figure}[t]
\centering
\includegraphics{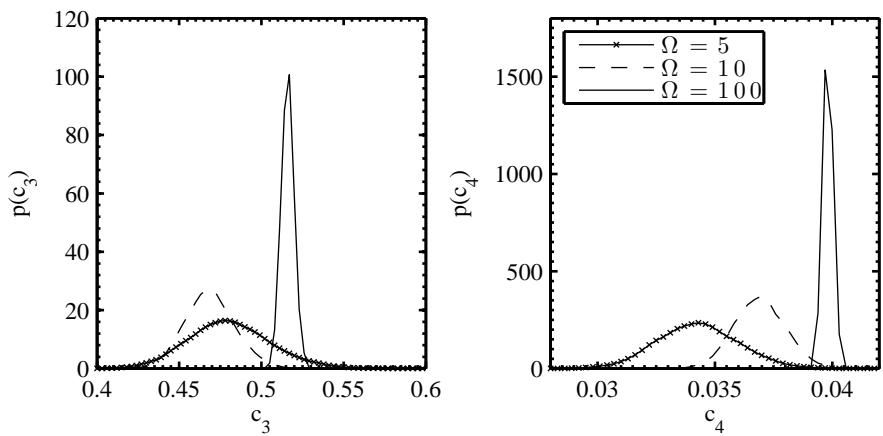}
\caption{Marginal posteriors for parameters $c_{3}$ (left panel) and $c_{4}$ right panel for different values of $\Omega$. Results are obtained by 10,000 posterior samples using RMHMC. \label{fig:dimer:marginal:omega}}
\end{figure}

Table~\ref{table:dimer:ESSOmega} compares the minimum Effective Sample Size (ESS) and the time normalised ESS obtained by all algorithms for different values of the system size parameter $\Omega$. The SMMALA and RMHMC samplers utilise the gradients and the Fisher Information of the approximate likelihood obtained by the LNA in order to make efficient proposals. As the system size increases and thus the LNA better approximates the true likelihood then mixing of the manifold MCMC algorithms improves. For this particular example we can see that good mixing can be achieved even for very small systems with only $\approx 25$ molecules, ($\Omega=5$). The M.H. sampler is not affected by the system size but its mixing is very poor in all cases. From the time normalised ESS we can also see that despite the improved mixing of RMHMC the computational cost is significant. On the contrary SMMALA provides a good tradeoff between mixing efficiency and computational cost. Finally, Table~\ref{table:dimer:MeanSDRMHMC} reports the marginal posterior means and standard deviations for different values of $\Omega$ obtained by RMHMC. The marginal posteriors for parameters $c_{3}$ and $c_{4}$ with $\Omega>=5$ are also shown in Figure~\ref{fig:dimer:marginal:omega}. Results from the MH and SMMALA samplers are similar and are omitted. For small system sizes we can observe that there is an increased bias of the Monte Carlo estimate while the posterior standard deviation is higher reflecting the high degree of uncertainty around the mean. The bias however significantly reduces as the system size increases and for $\Omega>= 5$ reasonable estimates can be obtained.

\begin{figure}[t]
\centering
\includegraphics{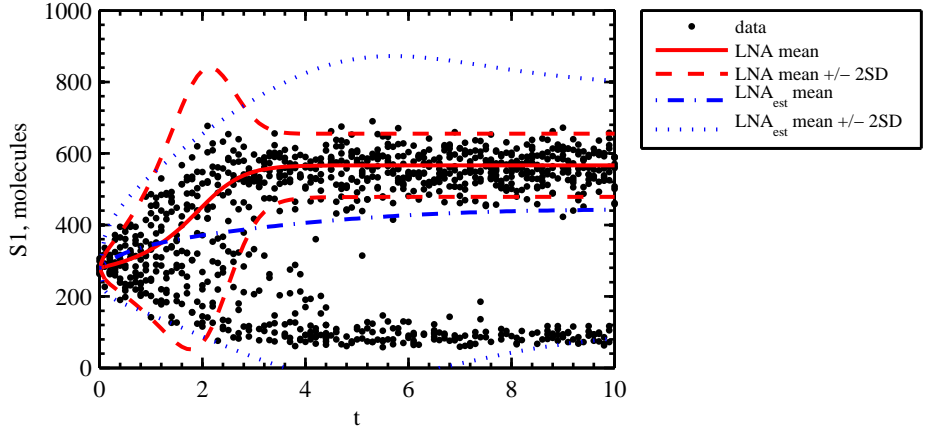}
\caption{Simulated time point data using SSA for the  Schl\"ogl reaction set and LNA predictions. Dots correspond to simulated data. The bold and dashed red lines correspond to the LNA prediction for the means and standard deviations using the true parameters. Doted blue lines correspond the LNA predictions using the posterior means for the rate parameters. (Online version in colour.)\label{fig:Schlogl:LNA}}
 \end{figure}

The second example from the chemical kinetics literature that we consider is the Schl\"ogl reaction set. 
\begin{eqnarray*}
R1 : 2S_{1} & \xrightarrow{c_{1}\Omega^{-1}} & 3S_{1} \\
R2: 3S_{1} & \xrightarrow{c_{2}\Omega^{-2}} & 2S_{1}\\ 
R3:\emptyset & \xrightarrow{c_{3}\Omega} & S_{1} \\
R4: S_{1} & \xrightarrow{c_{4}} & \emptyset .
\end{eqnarray*}
The corresponding state transition rates and state change matrix are given in equations~(\ref{eq:Schlogl:S}) and (\ref{eq:Schlogl:rateF}) respectively. The state of the system consists only of the number of molecules of a single species $\vect{X}(t) = S_{1}(t)$.
\begin{table}
\begin{minipage}[c]{0.4\linewidth}
\small
\centering
\begin{equation}
\vect{S} = \left(\begin{array}{cccc}1, & -1, & 1, & -1,\end{array}\right)
\label{eq:Schlogl:S}
\end{equation}
\end{minipage}
\hfill
\begin{minipage}[c]{0.6\linewidth}
\small
\centering
\begin {equation}
\vect{f}(\vect{X},\vect{\theta}) = \left(\begin{array}{l} 
						c_{1}\Omega^{-1}\frac{1}{2}S_{1}(S_{1}-1), \\
						c_{2}\Omega^{-2}\frac{1}{6}S_{1}(S_{1}-1)(S_{1}-2), \\
						c_{3}\Omega, \\
						c_{4}S_{1} \\
					         \end{array}\right)
\label{eq:Schlogl:rateF}
\end{equation}
\end{minipage}
\end{table}
The system is known to have two stable states which appear at different times depending on the size of the system. \cite{Wallace2012} have shown that the LNA fails to provide a reasonable approximation of this system even for large concentration numbers. Their numerical experiments demonstrate that the LNA approximation can only approximate one of the two modes depending on the initial conditions. Here our aim is to show that using the LNA to obtain an approximate posterior over the unknown reaction rate constants can be very misleading for bi-stable systems. Using the resulting posterior means for the reaction rates gives us an LNA that fails to approximate any of the two stable modes.

To demonstrate that we follow the same experimental procedure as in the previous example. That is, we simulate data using the SSA for the time interval $t_{i}\in[0,10]$, $t_{i}-t_{i-1}=0.1$ with fixed rate parameters and then use this data for posterior inference of the rate parameters using MCMC. Values for the true rate parameters and initial conditions where set as in \cite{Wallace2012}. Namely, $c_{1} = 0.003$, $c_{2}= 0.0001$, $c_{3} = 200$, $c_{4}=3.5$ and $\vect{X}(t_{0}) = 280\Omega$, where $\Omega$ was fixed to 1. After 10,000 burn-in samples all samplers converged to a posterior distributions with mean $$\mbox{E}_{p(\vect{\theta}|\vect{X})}[\vect{\theta}] \approx (0.130, 3.3e^{-4}, 3.5e^{+3}, 26.22)^{T}$$ and variance $$\mbox{var}_{p(\vect{\theta}|\vect{X})}[\vect{\theta}] \approx (1.2e^{-4}, 8.2e^{-10}, 8.6e^{+4}, 4.53)^{T}$$ 
The LNA obtained by using the posterior means for the rate constants is shown in Figure~\ref{fig:Schlogl:LNA} along with the data obtained by the SSA and the LNA using the true values for the rate constants. We can see that the LNA approximation obtained by the posterior means fails to approximate any of the two modes. Rather it approximates the empirical mean and variance of the data. 

\subsection{Single gene expression}\label{sec:signlegene}
Finally, to illustrate the applicability of the methodology to systems biology we also consider a simplified model for the biochemical reactions involved in the expression of a single gene to protein. The model presented in this section is the same with the model used in the study of \cite{Komorowski2009} and we adopt the same notation in order to make comparisons easier. Gene expression is modelled in terms of three biochemical species; DNA, mRNA and protein; and four chemical reactions or state transitions; transcription, mRNA degradation, translation and protein degradation. The model can be written in chemical reaction notation as
\begin{eqnarray*}
R1 : DNA & \xrightarrow{k_{R}(t)} & DNA + R \\
R2: R & \xrightarrow{\gamma_{R}} &  \emptyset\\ 
R3: R & \xrightarrow{k_{P}} & R+P \\
R4: P & \xrightarrow{\gamma_{P}} & \emptyset .
\end{eqnarray*}

The system state at time $t$ is $\vect{X}(t) = [R(t), P(t) ]^{T}$ where $R(t)$ and $P(t)$ are the number of mRNA and protein molecules respectively. The corresponding state dependent transition rates are $\vect{f}(\vect{X},t) = [ k_{R}(t), \gamma_{R}R(t), k_{P} R(t), \gamma_{P}P(t)]^{T}$ where $\gamma_{R}, k_{P}$ and $\gamma_{P}$ are unknown reaction rate constants. $k_{R}(t)$ is the time dependent transcription rate of the gene which for the purposes of this section is modelled as $$k_{R}(t) = b_{0}\exp(-b_{1}(t - b_{2})^{2})+b_{3}$$ where all the $b_{i}$s are also unknown parameters  controlling gene transcription. This corresponds to a transcription rate that due to some stimulus (experimental or environmental) increases for $t<b_{2}$ and then it drops towards the base line $b_{3}$ for $t>b_{2}$. Finally, the state change matrix for this set of reactions is given in equation~(\ref{eq:siglegene:S}).
\begin{equation}
\vect{S} = \left(\begin{array}{cccc} 1 & -1 & 0 & 0 \\0 & 0 & 1 & -1\end{array}\right).
\label{eq:siglegene:S}
\end{equation}

\begin{figure}[!t]
\begin{center}
\includegraphics[width=0.35\linewidth]{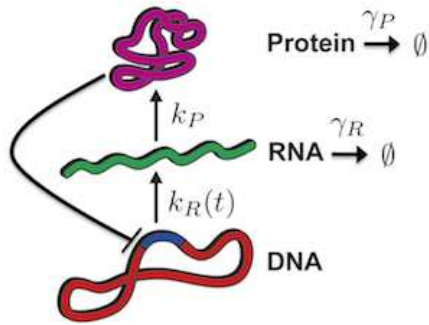}
\end{center}
\caption{Schematic representation of the auto-regulatory gene expression model with a negative feedback loop. A gene is transcribed into mRNA which is translated to a protein that suppresses gene transcription. (Online version in colour.)\label{fig:signlegene:schematic}}
 \end{figure}

As in the study of \cite{Komorowski2009} we also consider a non-linear extension of this model where the transcription rate of the gene $k_{R}(t)$ is a function of the protein concentration that the gene is transcribed to. This is modelled using a Hill function 
$$\hat{k}_{R}(t,P) = k_{R}(t)/(1+(P/H)^{n_{H}})$$
where for the experiments of this section we will set $H = b_{3}k_{P}/(2\gamma_{R}\gamma_{P})$ and $n_{H}=1/2$ making the protein an inhibitor of mRNA transcription. A schematic representation of this model is shown in Figure~\ref{fig:signlegene:schematic}. For the rest of this section we will refer to this model as the auto-regulatory single gene expression model.

\begin{figure}[!t]
\centering
\includegraphics{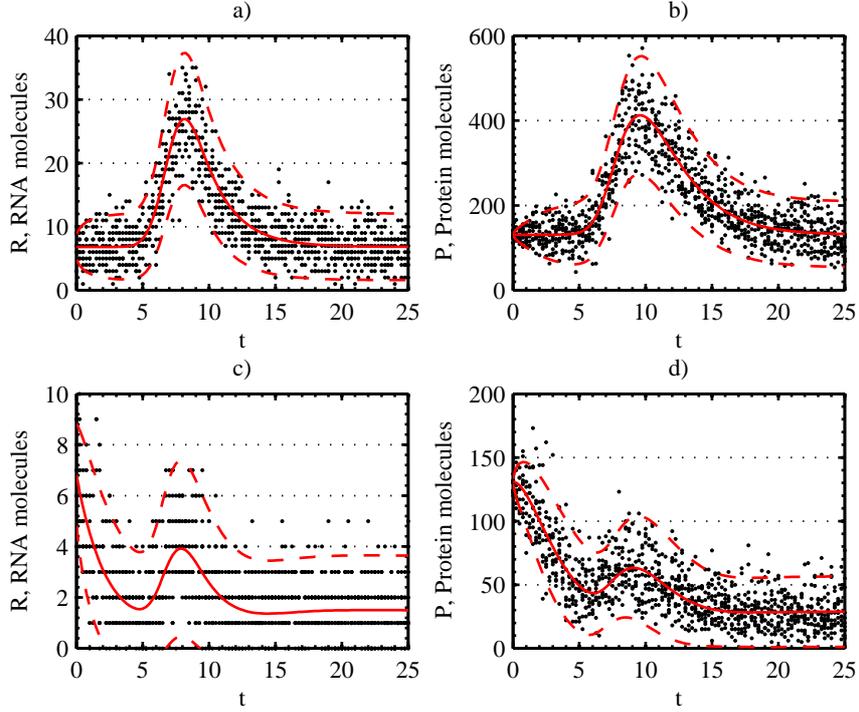}
\caption{Data simulated from the single gene expression model using SSA. Figures (a), (b), for the linear model and Figures (c), (d), for the auto-regulatory model. Dots correspond to 10 independent draws for each time point. The bold line is the mean predicted by LNA with the true model parameters and the dashed lines are the $+-2\times$ standard deviation predicted by LNA. Left column shows the mRNA molecules and right column the protein. (Online version in colour.)\label{fig:signlegene:traj}}
 \end{figure}

Using the transition probabilities $\vect{f}(\vect{X},t)$ and matrix $\vect{S}$ we simulate synthetic data using the Stochastic Simulation Algorithm (SSA) \cite{Gillespie2007} and sample at discrete time points. Values for the unknown rate constants and the parameters controlling gene transcription are shown in Table~\ref{table:signlegene:results}. The time interval is taken to be $t_{i} \in [0,25]$ while the interval between two observations $t_{i} - t_{i-1}= 0.25$. Each time point is sampled from an independent trajectory by starting the SSA from $t_{0}$ and simulate up to $t_{i}$  keeping only the state $\vect{X}(t_{i})$ and discarding the rest of the trajectory. This resembles the experimental conditions often encountered in biology where in order to make an observation the sample has to be ``sacrificed''. Finally for each time point we also generate 10 independent observations from different trajectories. Initial conditions $\vect{X}(t_{0})$ are simulated from a Poisson distribution with means $b_{3}/\gamma_{R}$ and $b_{3}k_{P}/(\gamma_{R}\gamma_{P})$ for the mRNA and protein molecules respectively. The system size parameter $\Omega$ is considered to be unknown and for this experiment is set to 1 such that concentrations are equal to the number of molecules. Figures~(\ref{fig:signlegene:traj}.a) and~(\ref{fig:signlegene:traj}.b) show data simulated from this process from the singe gene expression model as well as the LNA prediction. Simulated data for the auto-regulatory model are presented in Figures~(\ref{fig:signlegene:traj}.c) and~(\ref{fig:signlegene:traj}.d).

We use the simulated data to infer the unknown parameters $\vect{\theta} = (\gamma_{R}, k_{P},\gamma_{P}, b_{0}, b_{1}, b_{2}, b_{5} )^{T}$ by sampling using MCMC from the LNA approximate posterior
$$p(\vect{\theta}|\vect{X}) \propto p(\vect{X}|\vect{\theta})p(\vect{\theta}) = p(\vect{\theta})\prod_{i=1}^{N}\prod_{r=1}^{R}\Norm[\vect{X}_{r}(t_{i})|\vect{\phi}(t_i),\vect{V}(t_{i})]$$
where $r$ indexes independent samples for the same time point and $R=10$. 

\begin{figure}[!t]
\centering
\includegraphics{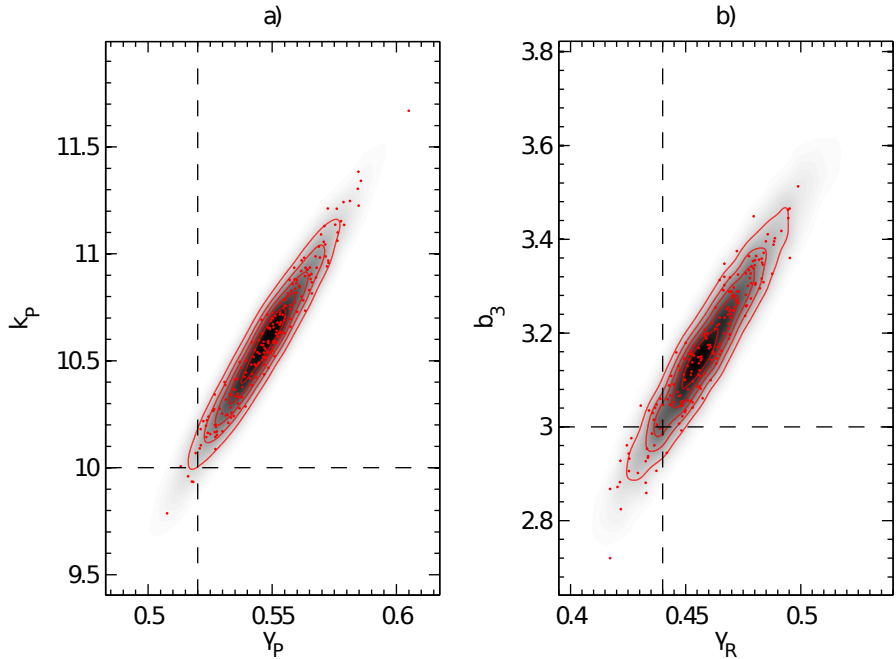}
\caption{Marginal joint posterior for parameters $\gamma_{P},k_{P}$ left panel, Figure (a), and $\gamma_{R},b_{3}$ right panel, Figure (b) for the single gene expression model. Dashed lines are the true values used to generate the synthetic data. Dots are samples from the posterior. Iso-contours and shaded region are obtained by kernel density estimation using posterior samples. (Online version in colour.) \label{fig:singlegene:marginal:post}}
 \end{figure}

\begin{figure}[t]
\centering
\includegraphics{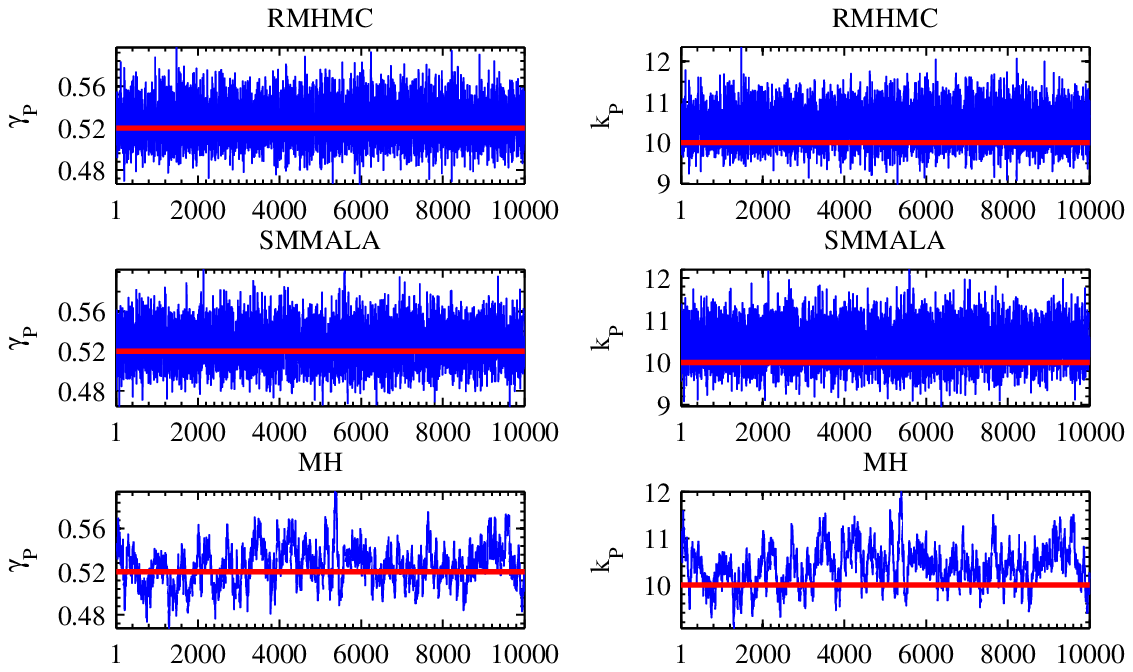}
\caption{Example trace plots from the auto-regulatory gene expression model for parameters $\gamma_{P}$ and $k_{P}$. Red solid line denotes the true values. (Online version in colour.) \label{fig:singlegene:trace}}
 \end{figure}

Table~\ref{table:signlegene:results} summarises the results from the MCMC chains for the two models of gene expression. Firstly, we can see that despite the relatively small number of molecules in both systems the LNA approximation provides very accurate estimates for the true parameters. Moreover we can see that the mixing of the Metropolis-Hastings sampler is very poor for both models while RMHMC and simplified Manifold MALA algorithms perform very well. This can be explained by the strong correlations between parameters in the posterior distribution preventing the M.H. sampler to make sufficiently large proposals. For example, the parameters $k_{P}, \gamma_{P}$ control mRNA translation and protein degradation respectively. The concentration of protein molecules is directly affected by the two rates and they are expected to be heavily correlated. In Figures~\ref{fig:singlegene:marginal:post}.a ~\ref{fig:singlegene:marginal:post}.b we show the marginal joint posterior for parameters $k_{P}, \gamma_{P}$ and $\gamma_{R}, b_{3}$ for the single gene expression model which exhibit very strong positive correlation. Finally figure~\ref{fig:singlegene:trace} compares the trace plots obtained from MH, SMMALA and RMHMC for parameters $\gamma_{P}$ and $k_{P}$ of the auto-regulatory gene expression model.

\begin{table}
\begin{center}
\begin{tabular}{lrrrrrrr}
\toprule
\multicolumn{8}{c}{Single gene expression model.} \\
\toprule
Parameters & $\gamma_{R}$ & $\gamma_{P}$ & $k_{P}$ & $b_{0}$ &$b_{1}$ & $b_{2}$ & $b_{3}$ \\
\midrule
True values & 0.44 & 0.52 & 10.0 & 15.0 & 0.40 & 7.0 & 3.0 \\
\midrule
\multicolumn{8}{c}{Metropolis-Hastings} \\
(A.R.) & (0.28) & (0.33) & (0.30) & (0.34) & (0.29) & (0.29) & (0.34) \\
($\epsilon$) & (0.013) & (0.007) & (0.008) & (0.022) & (0.056) & (0.007) & (0.016) \\
\midrule
Mean & 0.45 & 0.54 & 10.54 & 14.86 & 0.39 & 7.03 & 3.14\\
S.D. &  0.017 & 0.017 & 0.336 & 0.509 & 0.029 & 0.056 & 0.149\\
ESS & 42 & 34 & 34 & 149 & 117 & 58 & 44 \\
ESS/time & 1.42 & 1.15 & 1.15 & 5.05 & 3.96 & 1.96 & 1.49 \\
\midrule
\multicolumn{8}{c}{Simplified Manifold MALA (A.R.= 0.79, $\epsilon$ = 1.05 ) } \\
\midrule
Mean & 0.45 & 0.54 & 10.57 & 14.88 & 0.39 & 7.04 & 3.17\\
S.D. &  0.018 & 0.016 & 0.306 & 0.537 & 0.030 & 0.053 & 0.152\\
ESS & 2891 & 2911 & 2958 & 2787 & 3310 & 3183 & 2878 \\
ESS/time & 83.79 & 84.37 & 85.73 & 80.78 & 95.94 & 92.26 & 83.42 \\
\midrule
\multicolumn{8}{c}{Manifold HMC (A.R.= 0.84, $\epsilon$ = 0.91,  L=5 ) } \\
\midrule
Mean & 0.46 & 0.54 & 10.57 & 14.95 & 0.39 & 7.04 & 3.18\\
S.D. &  0.018 & 0.015 & 0.300 & 0.555 & 0.030 & 0.052 & 0.153\\
ESS & 7731 & 8238 & 8304 & 7160 & 7380 & 7791 & 7950 \\
ESS/time & 0.52 & 0.55 & 0.56 & 0.48 & 0.49 & 0.52 & 0.53 \\
\bottomrule
\multicolumn{8}{c}{Auto-regulatory single gene expression model.} \\
\bottomrule
\multicolumn{8}{c}{Metropolis-Hastings} \\
(A.R.) & (0.26) & (0.36) & (0.31) & (0.33) & (0.24) & (0.30) & (0.35) \\
($\epsilon$) & (0.028) & (0.012) & (0.016) & (0.071) & (0.231) & (0.019) & (0.029) \\
\midrule
Mean & 0.4360 & 0.52 & 10.40& 14.61 & 0.40 & 6.82 & 3.13\\
S.D. &  0.016 & 0.018 & 0.424 & 1.089 & 0.076 & 0.090 & 0.142\\
ESS & 201 & 71 & 73 & 465 & 339 & 420 & 239 \\
ESS/time & 6.12 & 2.16 & 2.22 & 14.17 & 10.33 & 12.80 & 7.28 \\
\midrule
\multicolumn{8}{c}{Simplified Manifold MALA (A.R.= 0.71, $\epsilon$ = 1.17) } \\
\midrule
Mean & 0.43 & 0.52 & 10.44 & 14.24 & 0.38 & 6.82 & 3.12\\
S.D. &  0.016 & 0.018 & 0.422 & 1.125 & 0.075 & 0.091 & 0.142\\
ESS & 2990 & 3270 & 3454 & 3124 & 3164 & 3316 & 3195 \\
ESS/time & 76.86 & 84.06 & 88.79 & 80.30 & 81.33 & 85.24 & 82.13 \\
\midrule
\multicolumn{8}{c}{Manifold HMC (A.R.= 0.82, $\epsilon$ = 0.91,  L=5 ) } \\
\midrule
Mean & 0.43 & 0.52 & 10.43 & 14.52 & 0.40 & 6.82 & 3.13\\
S.D. &  0.016 & 0.017 & 0.412 & 1.158 & 0.078 & 0.089 & 0.144\\
ESS & 6532 & 6593 & 6614 & 5112 & 5384 & 6595 & 6642 \\
ESS/time & 0.41 & 0.41 & 0.41 & 0.32 & 0.34 & 0.41 & 0.42 \\
\bottomrule
\end{tabular}
\caption{Marginal posterior means and standard deviations for the parameters of the single gene expression model using simulated data. The ESS is calculated for chains of 10,000 samples after a burn-in period of 10,000 iterations with initial parameters randomly sampled from the prior. Average acceptance rate (A.R.) and sampler parameters are shown in parenthesis. Notice that for the Metropolis-Hastings sampler a different proposal is used for each parameter. The prior for all parameters was $\log_{10}\Norm(0.0,2.0)$. \label{table:signlegene:results}}
\end{center}
\end{table}

\section{Conclusions and Future Work}
\label{sec:conclusions}
Bayesian inference for Markov jump processes is a challenging problem which has many important practical applications. Previous research  \cite{Boys2008} has shown that although exact inference is possible, the computational cost and the autocorrelation of the Markov chains is such that limits its applicability to small problems. The main problem stems from the requirement to simulate the MJP for the trajectory of the system between discrete observations. \cite{Golightly2011} has shown that by considering a diffusion approximation the simulation can be performed in a much more efficient manner. In this paper we considered the linear noise approximation which only requires to simulate a system of ordinary differential equations while the stochastic fluctuations have an exact analytic solution. The linear noise approximation is valid only when the system is sufficiently close to its thermodynamic limit, a condition that is also required for the diffusion approximation. Previous research on the linear noise approximation \cite{Komorowski2009} has focussed on the Metropolis-Hastings sampler. We have demonstrated here that when the posterior distribution exhibits strong correlation between parameters then the Metropolis-Hastings sampler has strong auto-correlations. Such correlations are very common for chemical reaction and gene regulatory systems. The Riemann manifold MCMC algorithms we considered in this work exploit the geometric structure of the target posterior in order to design efficient proposal mechanisms. In particular the simplified Manifold MALA algorithm is a conceptually simple algorithm which provides a good trade-off between computational cost and sample auto-correlation.

Although the problems considered in this work are relatively small, but certainly non-trivial, we believe that the proposed methodology is applicable for larger and more complex systems. The systems we studied in this paper all have a linear dependence on the unknown parameters and we have not observed any local modes in our simulations. The analysis of such systems is the subject of on-going work. Moreover, in real applications it is not possible to observe the populations of all species and there is an additional measurement error term. Extension of the LNA to handle such cases is straight forward, see \cite{Komorowski2009} for example, however the effect of partial observations and measurement error on the MCMC inference is something that needs to be studied in more detail. 



\end{document}